\documentclass{article}

\usepackage{microtype}
\usepackage{graphicx}
\usepackage{booktabs} 

\usepackage{hyperref}

\usepackage{amsmath,amssymb}
\usepackage{cleveref}
\usepackage{url}
\usepackage{soul}
\usepackage{multirow}
\usepackage[toc,page]{appendix}
\usepackage{subcaption}
\usepackage{blkarray, bigstrut}
\usepackage{listings}
\usepackage{pgfplots}

\usepackage[accepted]{icml2022_noi}

\icmltitlerunning{Scaling up Ranking under Constraints for Live Recommendations}

\begin{document}

\twocolumn[
\icmltitle{Scaling up Ranking under Constraints for Live Recommendations\\by Replacing Optimization with Prediction}




\begin{icmlauthorlist}
\icmlauthor{Yegor Tkachenko}{cu}
\icmlauthor{Wassim Dhaouadi}{su}
\icmlauthor{Kamel Jedidi}{cu}
\end{icmlauthorlist}

\icmlaffiliation{cu}{Columbia University}
\icmlaffiliation{su}{Stanford University}

\icmlcorrespondingauthor{Yegor Tkachenko}{ytkachenko21@gsb.columbia.edu}

\icmlkeywords{Scalable ranking under constraints, recommender systems, ethical decision making}

\vskip 0.3in
]

\printAffiliationsAndNotice{} 
\begin{abstract}
Many important multiple-objective decision problems can be cast within the framework of ranking under constraints and solved via a weighted bipartite matching linear program. Some of these optimization problems, such as personalized content recommendations, may need to be solved in real time and thus must comply with strict time requirements to prevent the perception of latency by consumers. Classical linear programming is too computationally inefficient for such settings. We propose a novel approach to scale up ranking under constraints by replacing the weighted bipartite matching optimization with a prediction problem in the algorithm deployment stage. We show empirically that the proposed approximate solution to the ranking problem leads to a major reduction in required computing resources without much sacrifice in constraint compliance and achieved utility, allowing us to solve larger constrained ranking problems real-time, within the required 50 milliseconds, than previously reported.
\end{abstract}

\section{Introduction}

Many important multiple-objective decision problems can be cast within the framework of ranking under constraints. Content delivery platforms, such as TikTok, Pinterest, Instagram, Facebook news feed, and Google search engine, are prominent examples. Such platforms constantly have to select and prioritize content from a large dynamic library to show within the restrictions of available display space and consumer's attention span, across millions of users \citep{ansari2000internet,covington2016deep,mnih2008probabilistic}. The recommendations need to balance multiple objectives -- the primary objective of optimizing revenue or its proxy, as well as auxiliary objectives, such as diversity, ethics, recency, and other properties of the recommended content \citep{celis2017ranking,white2000requirement}. To avoid the perception of latency, ranking must be returned in less than 100 milliseconds \citep{miller1968response}, which, given the time required to send and receive information online, means ranking has to be computed within 50 milliseconds \citep{zhernov2020nodehopper}. 

The problem of ranking under constraints in recommender settings can be formulated as a weighted bipartite matching linear program \citep{singh2018fairness,biega2018equity}. The formulation can handle varied constraints and flexible preferences over item-rank combinations. However, classical linear programming is too computationally inefficient to meet the latency requirement for live deployment. 

We address this problem of speed. We propose a scalable algorithm for ranking under constraints based on statistical sampling and a dual formulation of the weighted constrained bipartite matching program \citep{shah2017online,mehta2012online}. In particular, we propose to replace the problem of online optimization with a prediction problem. We solve the dual program for optimal shadow prices on a sample of users in offline settings, where speed is not critical. We then train a model to predict users' optimal shadow prices from users' characteristics. In online settings, the model can predict the shadow prices directly based on observed user covariates, without having to solve the time-consuming optimization problem, allowing us to quickly compute the ranking for any such user. We demonstrate empirically that the proposed approximate solution to the ranking problem leads to a substantial reduction in required computing resources without much sacrifice in constraint compliance and achieved utility, allowing us to solve larger constrained ranking problems real-time, within the required 50 milliseconds, than previously reported \citep{zhernov2020nodehopper}. The code to reproduce the results in this paper is made available on GitHub.\footnote{\scriptsize \url{https://github.com/computationalmarketing/scalable_ranking_under_constraints}}

{\bfseries Significance\ } Our fast constrained ranking framework can empower ethical personalization at scale across online content platforms. It could also tackle time-sensitive assignment and matching problems beyond content recommendations, e.g., robot swarm task allocation. Further, our results show the power of the idea of predicting optimization solutions, which could find many uses in AI research and practice.

\section{Optimization framework}

Consider an example where a recommender system ranks $m$ items (e.g., movies) to show to a given user. Assume that the decision maker's utility is maximized by showing items in the order that maximizes user's utility -- so decision maker's utility mirrors user's utility. This is in line with probability ranking principle, which states that, for optimal retrieval, documents should be ranked in order of the probability of relevance or usefulness \citep{robertson1977probability}. Let $U$ be an $m\times m$ utility matrix that captures decision maker's preferences. $U_{ij}$ gives decision maker's utility from assigning item $i$ to position $j$ in the ranking. In other words, each row in this matrix represents scores that item $i$ would get by being placed in different ranking positions $j$. A common assumption here is that the utility of an item should decline / be discounted with the lower position in the ranking because the user is less likely to be exposed to an item when it is displayed lower in the list \citep{singh2018fairness}. 

\subsection{Optimal assignment}
An assignment of items to positions in the ranking can be captured by an $m\times m$ permutation matrix $P\in\mathcal{P}$ ($\mathcal{P}$ is the space of permutation matrices): $P_{ij}=1$ if item $i$ is assigned to position $j$, $P_{ij}=0$ otherwise (each item is assigned to a single rank, and each rank holds a single item; in other words, all rows and columns sum to one; $P^T\mathbf{1}=\mathbf{1}$ and $P\mathbf{1}=\mathbf{1}$). We can write down the utility from a given assignment as a sum over element-wise product between $U$ and $P$: $\sum_i\sum_j U_{ij}P_{ij}$. This sum is equal to the trace of the dot product of the two matrices $\textbf{tr}(U^TP)=\sum_i\sum_j U_{ij}P_{ij}$. The goal is to find $P$ that maximizes $\textbf{tr}(U^TP)$. More formally, we have the following optimization problem, also known as the weighted bipartite matching problem or the assignment problem:
\begin{equation}
\label{eq:primal_unc}
\begin{split}
\max_{P}&\ \textbf{tr}(U^TP)\\
&\ P\in\mathcal{P}\text{.}
\end{split}
\end{equation}
The Hungarian algorithm gives an optimal solution \citep{edmonds1972theoretical,tomizawa1971some,kuhn1955hungarian,kuhn1956variants}.

\subsection{Adding constraints}

Ranking under constraints problem can be formulated as an instance of the constrained bipartite graph matching program, following \citet{singh2018fairness,biega2018equity}. Classical formulation of the problem is as follows:
\begin{equation}
\label{eq:primal}
\begin{split}
\max_{P}&\ \textbf{tr}(U^TP)\\
s.t.&\ \textbf{tr}(A_k^TP) \geq b_k\ \forall \ k\in1:K\\
&\ P\in\mathcal{P}\text{.}
\end{split}
\end{equation}
This program (\cref{eq:primal}) is different from the previous one (\cref{eq:primal_unc}) only in that $K$ constraints have been added: an $m\times m $ matrix $A_k$ together with a scalar $b_k$ track if $P$ satisfies a $k$-th constraint. $A_k$ can be interpreted as an auxiliary utility matrix corresponding to an alternative objective. The formulation can accommodate a variety of constraints, such as quotas by item type (e.g., protected category), constraints on inverse-rank-weighted average of item characteristics (e.g., release recency), etc. 
Addition of the constraints, while $P$ is restricted to be an integer permutation matrix, renders this combinatorial optimization problem intractable.

\subsection{Primal program}
To proceed, we remove the integer requirement, constraining $P$ to be a doubly stochastic matrix.
\begin{equation}
\label{eq:primal_rel}
\begin{split}
\max_{P}&\ \textbf{tr}(U^TP)\\
s.t.&\ \textbf{tr}(A_k^TP) \geq b_k\ \forall \ k\in1:K\\
&\ P\in\mathbb{R}^{m\times m},\ P\textbf{1} = \textbf{1},\ P^T\textbf{1} = \textbf{1},\ P\geq 0\text{.}
\end{split}
\end{equation}
This convex relaxation of the integer problem (\cref{eq:primal_rel}) can be solved using linear programming \citep{bubeck2014convex}. The solution $P$ may be fractional, but integer solutions can be obtained via Birkhoff-von Neumann decomposition \citep{birkhoff1940lattice}. We refer to this formulation as the {\it primal} program. 

\subsection{Dual program}
\label{sec:sample}

We also consider the dual formulation of the weighted bipartite graph matching problem \citep{roth1993stable,shah2017online,mehta2012online}, obtained by application of the duality theorem \citep{boyd2004convex} -- the {\it dual} program:
\begin{equation}
\label{eq:dual}
\begin{split}
\max_{\lambda,\alpha,\beta}&\ \lambda^T b + \alpha^T\mathbf{1} + \beta^T \mathbf{1}\\
s.t.&\ U + \sum_k\lambda_{k} A_{k} + \mathbf{1}\alpha^T + \beta \mathbf{1}^T\leq 0\\
&\ \lambda\geq 0\text{.}
\end{split}
\end{equation}
Here $\lambda_{k}$ is a scalar shadow price corresponding to constraint $k$ ($\lambda$ is a vector of all $K$ shadow prices). Shadow prices have rich interpretation. For example, when constraint $k$ specifies the minimum number of items from a protected category among top $n$ items, $\lambda_{k}$ can be interpreted as a `boost' in terms of utility that has to be applied to protected category items for them to be shifted up in the ranking to satisfy constraint $k$. Thus, shadow prices $\lambda$ can be informative about decision maker's preferences and the cost of the constraint. Shadow prices $\alpha$ and $\beta$ correspond, respectively, to column and row sum constraints on $P$. 

The optimized permutation ranking matrix can be obtained by applying the Hungarian algorithm to the modified utility matrix $S=U+\sum_k\lambda_{k} A_{k}$ for a given set of items \citep{shah2017online}. This suggests another view of $\lambda$ as optimal weights in a linear combination of primary ($U$) and auxiliary ($A_k$) utility matrices -- such that the resulting modified utility matrix $S$ directly yields optimal assignment upon application of the Hungarian algorithm. 
Thus, while the primary program searches directly for doubly stochastic matrix $P$ capturing optimal assignment, the dual program searches for $\lambda$ capturing the optimal combination of utility matrices $U$ and $A_k$ behind multiple objectives to yield the optimal assignment indirectly. The latter method proves to be useful for speeding up the computation.

\section{Speeding up ranking under constraints}

Off-the-shelf solvers struggle to solve the above linear programs in real time (under 50 milliseconds) in general. For large numbers of items and constraints ($\geq 500$ ranked objects and $\geq 5$ constraints), even custom solvers relying on special problem structure fail \citep{zhernov2020nodehopper}.

\subsection{Prediction instead of optimization}
To address computational inefficiencies of linear programming, we propose to replace real-time optimization with prediction. 
We first solve the dual program to obtain optimal shadow prices on a sample of users offline, where speed is not critical. We can then train a supervised learning model to predict users' optimal shadow prices from users' characteristics. In online settings, for a new user we have not seen before, the model can predict the optimal shadow prices $\hat\lambda$ based on observed covariates, without solving the time-consuming optimization problem. We can then directly compute the adjusted utility matrix $\hat S=U+\sum_k\hat\lambda_{k} A_{k}$ for the new user and obtain the optimal permutation matrix from $\hat S$ afterwards. \Cref{algo:b} presents our proposed procedure.

\begin{algorithm}[tb]
    \caption{Scalable personalized constrained ranking}
    \label{algo:b}
\begin{algorithmic}
\STATE {\bfseries Input:} For each ranking instance $l$ (user), utility matrix $U^{(l)}$, constraint arrays $A_k^{(l)}$, $b_k^{(l)}$, and covariates $X^{(l)}$.
\FOR{$l$ {\bfseries in} \textit{Offline Train Set}}
    \STATE Solve dual linear prog. (\cref{eq:dual}) for $U^{(l)}$, $A_k^{(l)}$, and $b_k^{(l)}$.
    \STATE Store optimal $\lambda^{(l)}$.
\ENDFOR
\STATE Train prediction model $f(X_{train})\rightarrow \lambda_{train}$ based on \textit{Offline Train Set}.
\FOR{$l$ {\bfseries in} \textit{Online Test Set}}
    \STATE Predict $\hat\lambda^{(l)}\leftarrow f(X^{(l)})$.
    \STATE Compute $\hat S^{(l)} = U^{(l)} + \sum_k\hat\lambda_k^{(l)}A_k^{(l)}$.
    \STATE Determine an optimal assignment $P^{(l)}$ based on $\hat S^{(l)}$ via Hungarian method, greedy $1/2$-approximation algorithm, via identity permutation if $\hat S^{(l)}$ is inverse Monge, or via sorting + identity permutation in case of special permuted inverse Monge structure. Alternatively, an optimal assignment could be approximated with a neural net prediction.
\ENDFOR
\end{algorithmic}
\end{algorithm}

{\bfseries Layers of approximation\ } Note that all of the discussed practical constrained optimization formulations only provide approximate solutions, first and foremost stemming from the convex relaxation. In case of the primal problem relaxation in \cref{eq:primal_rel}, approximation arises because we need to move from doubly-stochastic matrix to an integral permutation matrix via Birkhoff-von Neumann decomposition. Finding such a decomposition with minimum number of terms is an NP-hard problem in itself and admits only heuristic solutions, with the commonly used greedy heuristic having complexity $O(m^2)$ \citep{duff2001algorithms,dufosse2016notes}. It results in, at most, $m^2$ distinct permutation matrices, which can then be sampled based on this decomposition. Constraints are satisfied only asymptotically. In case of the dual formulation in \cref{eq:dual}, the source of approximation arises because the Hungarian algorithm run on $S$ returns one of the valid extremal solutions from Birkhoff polytope \citep{schrijver2003combinatorial}, and thus also satisfies primal constraints asymptotically. That is, the use of shadow prices $\lambda$ obtained as a solution to \cref{eq:dual} to compute $S=U+\sum_k\lambda_{k} A_{k}$ can result in multiple constraint-compliant and constraint-non-compliant assignments that are tied in utility. In practice, we can often break ties heuristically in favor of auxiliary objectives (constraint compliance) by computing $S=U+\sum_k(1+\varepsilon) \lambda_{k} A_{k}$, where $\varepsilon > 0$ is a tuning parameter. \Cref{fig:ex2} gives an example of solving the dual program and handling the case of multiple solutions. 

The statistical estimation of $\lambda$ adds an additional layer of approximation, however, simple prediction algorithms can give us good estimates under mild assumptions. For example, a k-nearest neighbor regressor yields a consistent estimate of $E[\lambda|X=x]$, where $X$ denotes user covariates, given a large number of i.i.d. observations, a large enough number of neighbors used in the prediction, and some technical conditions \citep{devroye1994strong}. Intuitively, with enough data, we can predict $\lambda$ arbitrarily well, limited only by the expressive power of covariates and systematic noise.

Taking note of these technical issues, we show empirically in \Cref{sec:empirical} that we can consistently obtain constraint-compliant solutions with minimal loss of optimality.

\begin{figure*}[h] \footnotesize
\[\begin{split}
  U=
  \begin{blockarray}{*{4}{c} l}
    \begin{block}{*{4}{>{$\footnotesize}c<{$}} l}
      R1 & R2 & R3 & R4 & \\
    \end{block}
    \begin{block}{[*{4}{c}]>{$\footnotesize}l<{$}}
      5 & 4 & 2 & 1 \bigstrut[t]& I1 \\
      5 & 3 & 3 & 2 & I2 \\
      3 & 3 & 3 & 3 & I3 \\
      2 & 1 & 0 & 0 & I4 \\
    \end{block}
  \end{blockarray}
 \end{split}
 \quad\quad\quad\quad
 \begin{split}
  A_1=
  \begin{blockarray}{*{4}{c} l}
    \begin{block}{*{4}{>{$\footnotesize}c<{$}} l}
      R1 & R2 & R3 & R4 & \\
    \end{block}
    \begin{block}{[*{4}{c}]>{$\footnotesize}l<{$}}
      0 & 0 & 0 & 0 \bigstrut[t]& I1 \\
      0 & 0 & 0 & 0 & I2 \\
      1 & 0.6 & 0.5 & 0.4 & I3 \\
      0 & 0 & 0 & 0 & I4 \\
    \end{block}
  \end{blockarray}
 \end{split}
 \quad\quad\quad\quad
  \begin{split}
   \textbf{tr}(A_1^TP)\geq 0.7
 \end{split}
 \]
 \\
 \[  
 \begin{split}
   P=
  \begin{blockarray}{*{4}{c} l}
    \begin{block}{*{4}{>{$\footnotesize}c<{$}} l}
      R1 & R2 & R3 & R4 & \\
    \end{block}
    \begin{block}{[*{4}{c}]>{$\footnotesize}l<{$}}
      0 & 1 & 0 & 0 \bigstrut[t]& I1 \\
      0 & 0 & 1 & 0 & I2 \\
      1 & 0 & 0 & 0 & I3 \\
      0 & 0 & 0 & 1 & I4 \\
    \end{block}
  \end{blockarray}
 \end{split}
 \quad\quad\quad\quad
   \begin{split}
   \lambda_1=4.0
 \end{split}
 \quad\quad\quad\quad
 S = U + \lambda_1A_1 =   
 \begin{blockarray}{*{4}{c} l}
    \begin{block}{*{4}{>{$\footnotesize}c<{$}} l}
      R1 & R2 & R3 & R4 & \\
    \end{block}
    \begin{block}{[*{4}{c}]>{$\footnotesize}l<{$}}
      5 & 4 & 2 & 1 \bigstrut[t]& I1 \\
      5 & 3 & 3 & 2 & I2 \\
      7 & 5.4 & 5 & 4.6 & I3 \\
      2 & 1 & 0 & 0 & I4 \\
    \end{block}
  \end{blockarray}
 \]
\caption{\textit{Problem formulation example.} $U$ is a utility matrix. $U_{ij}$ gives utility a decision maker obtains from placing item $i$ in rank position $j$ (e.g., as predicted by a recommender system). $P$ is a permutation matrix. $P_{ij}=1$ if item $i$ is assigned to position $j$, and $P_{ij}=0$ otherwise. 
We want to find an optimal assignment, subject to constraint $\textbf{tr}(A_1^TP)\geq 0.7$. Here $A_1$ is an auxiliary utility matrix indicating items with some valued attribute, the utility decaying as item gets ranked lower. The constraint sets the minimum acceptable level of auxiliary utility. The only way to satisfy the constraint here is to place item 3 in rank 1. In more complex cases, we have to resort to formal optimization to solve the dual program. $\lambda_1=4$ is the shadow price returned by the optimization program and $S=U+\lambda_1A_1$ is the adjusted utility matrix that we can find an optimal assignment from. Critically, there is a tie here. Presented $P$ gives the optimal assignment satisfying the constraint, with $\textbf{tr}(S^TP)=14$. However, a different $(i,j)$ assignment $(2,1),(1,2),(3,3),(4,4)$ gives the same total adjusted utility but violates the constraint. This shows that a dual program solution need not be unique and only satisfies constraints asymptotically. Fortunately, we can often heuristically break the tie in favor of the constraint-compliant solution by computing $S=U+\sum_k(1+\varepsilon) \lambda_{k} A_{k}$, where $\varepsilon>0$ (e.g., $\varepsilon=10^{-4}$). We treat $\varepsilon$ as a tuning parameter.}
 \label{fig:ex2}
\end{figure*}

\subsection{Special problem structure for further speed up}
\label{sec:monge}

An optimal assignment permutation matrix $P$ ($m\times m$) can be computed from $S$ in the most general settings via the Hungarian algorithm with worst-case run-time complexity $O(m^3)$ \citep{edmonds1972theoretical,tomizawa1971some,bougleux2017hungarian}. 
If $m$ is large, this computational step can become a bottleneck and so we would like to find a faster algorithm. One approach that has been proposed is to use a greedy $1/2$-approximation algorithm \citep{avis1983survey,preis1999linear,eff_alg} with lower worst-case complexity $O(m^2)$. However, the weight of its matching can be as low as $1/2$ of the maximum weight computed by the Hungarian algorithm.
Luckily, we can get the best of both worlds in terms of speed and guaranteed matching optimality if we restrict $S$ to possess a special structure. 

\subsubsection{Fixed discounting and rearrangement inequalities}
Consider a `fixed discounting' formulation $U=u\gamma^T$ and $A_k = a_k\gamma^T$ for all $k$, with $\gamma,u,a_k\in \mathbb{R}^m$. Vectors $u$ and $a_k$ give primary and auxiliary utilities of items, independent of assigned rank. $\gamma$ is a discount vector along the ranks. Then $S=U+\sum_k\lambda_kA_k=(u+\sum_k\lambda_ka_k)\gamma^T=s\gamma^T$, where $s=u+\sum_k\lambda_ka_k$. We seek a permutation matrix $P\in\mathcal{P}$ that maximizes $\textbf{tr}(S^TP)$. Intuitively, the lower the item's rank, the lower is its observation probability, so $u$ and $a_k$ utilities should be discounted proportionately \citep{singh2018fairness}.\footnote{For example, discounted cumulative gain (DCG) approach expresses discounting factor as $\gamma_i = 1/\log_2(1+i)$ with $i\in1:m$, which is non-increasing \citep{jarvelin2002cumulated,singh2018fairness,biega2018equity}. Simple discounting $\gamma_i = d^i$ for $d\in(0,1]$ also fits the profile.}

Let parenthesis indexing denote descending ordering of the sequences so that $s_{(1)} \geq s_{(2)} \geq \cdots \geq s_{(m)}$ and $\gamma_{(1)} \geq \gamma_{(2)} \geq \cdots \geq \gamma_{(m)}$, for $s,\gamma$ that are in arbitrary order. \citet{hardy1952} have proven the following {\it rearrangement inequality} applicable to our case: 
\begin{equation}
\label{eq:rear}
\sum_{i=1}^ms_{(i)}\gamma_{(i)} \geq \sum_{i=1}^ms_{i}\gamma_{(i)} \geq \sum_{i=1}^ms_{(m-i+1)}\gamma_{(i)}\text{.}
\end{equation}
It follows immediately from this inequality that when two sequences $s$ and $\gamma$ are sorted in a descending order, then identity permutation matrix $P=I_{m\times m}$, from among all permutation matrices, maximizes $s^TP\gamma=\textbf{tr}(\gamma s^TP)$. 

Discounting vector $\gamma$ is determined by the decision maker and it makes intuitive sense to use discounting that is in the descending order (we will also assume $\gamma>0$). However, $s = u+\sum_k\lambda_ka_k$ need not be in the descending order. To handle this, define $\text{argsort}(x)=R_x\in\mathcal{P}$ such that $R_xx$ is in descending order. Once we have obtained $S = (u+\sum_k\lambda_ka_k)\gamma^T$, we can look at its first column $S[:,1]=\gamma_1(u+\sum_k\lambda_ka_k)$. If $\gamma_1>0$, $R_s=\text{argsort}(S[:,1])=\text{argsort}(u+\sum_k\lambda_ka_k)$. Then $R_sS=\text{sort}(s)\gamma^T$ and, by rearrangement inequality, $I_{m\times m} = \text{argmax}_{P\in\mathcal{P}} \textbf{tr}(S^TR_s^TP)$. It follows that $R_s^TI_{m\times m}= R_s^T = \text{argmax}_{P\in\mathcal{P}} \textbf{tr}(S^TP)$. Thus, if $\gamma>0$ and is in descending order, we only need to sort $S$ on its first column to recover optimal assignment via identity permutation. This gives us an $O(m \log m)$ time complexity algorithm, in contrast to $O(m^3)$ worst time complexity of the Hungarian algorithm and $O(m^2)$ worst time complexity of the greedy algorithm, making this method more scalable and the intentional restriction on the decision problem with fixed discounting an attractive approach.

The presented optimality proof via rearrangement inequality explains the great performance of the greedy sorting algorithm observed (but left without an explanation) by \citet{shah2017online} in the context of their Reference CTR model, which is a form of fixed discounting model presented above. Notably, the same fixed discounting formulation appears in a different context as a necessary condition for operation of the custom efficient solver by \citet{zhernov2020nodehopper}.

\subsubsection{Generalization and alternatives} Monge property is a generalization of rearrangement inequalities \citep{holstermann2017generalization}. If $S$ is (weak) inverse Monge, then optimal permutation matrix is an identity matrix \citep{burkard1996perspectives}. If $S$ is inverse Monge after sorting, it is called permuted Monge \citep{hutter2020estimation}. Fixed discounting problem structure yields Monge structure -- for $s$ and $\gamma$ non-increasing, $S = s\gamma^T$ is inverse Monge \citep{burkard2007monge}. However, $S$ can be inverse Monge without the discounting structure, as Monge condition is more general. If Monge condition is not satisfied, we can use the greedy $1/2$-approximation algorithm \citep{avis1983survey,preis1999linear,eff_alg} with worst-case complexity $O(m^2)$ before using the $O(m^3)$ Hungarian algorithm. There is also research on using neural nets to approximate the Hungarian algorithm optimal assignment with complexity $O(m^2)$ or $O(m)$ (depending on the neural net type) \citep{lee2018deep}. See \Cref{appendix:a} for an extended discussion.

\subsection{Unbalanced case} To simplify exposition, we have presented the results for square $U$ and $A_k$ matrices. Unbalanced case (rectangular $U$ and $A_k$, $m_1$ items and $m_2$ rank positions, $m_1>m_2$ -- some items remain unassigned) can also be accommodated. The dual program can be augmented to tackle a different number of vertices in a bipartite graph, relaxing the restriction on the permutation matrix from equality to inequality \citep{mehta2012online} (see \Cref{eq:unbalanced}). 
Optimal assignment can be found based on rectangular $S_{m_1\times m_2}$. The Hungarian algorithm can be applied with complexity $O(m_1m_2^2)$ \citep{bougleux2017hungarian}. Greedy $1/2$-approximation algorithm can be applied with complexity $O(m_1m_2)$ \citep{preis1999linear}. In the fixed discounting formulation, we can apply the algorithm without change, taking elements from the main ($i=j$) diagonal after sorting $S$ on the first column with worst time complexity $O(m_1\log m_1)$, as discussed earlier. In case of general unbalanced inverse Monge matrices, \citet{vaidyanathan2013faster} proposed a shortest path $O(m_1m_2)$ algorithm. \citet{burkard1996perspectives} mention an alternative algorithm based on dynamic programming with complexity $O((m_1-m_2+1)m_2)$, but do not provide implementation details. Optimal assignment algorithms for special types of unbalanced Monge matrices have been studied by \citet{aggarwal1992efficient}.

\begin{figure*} \footnotesize
\begin{equation}
\label{eq:unbalanced}
\begin{split}
&\textbf{Unbalanced primal}\\
\max_{P}&\ \textbf{tr}(U^TP)\\
s.t.&\ \textbf{tr}(A_k^TP) \geq b_k\ \forall \ k\in1:K\\
&\ P\in\mathbb{R}^{m_1\times m_2},\ P\textbf{1}_{m_2} \leq 1,\ P^T\textbf{1}_{m_1} = 1,\ P\geq 0
\end{split}
\quad\Rightarrow\quad
\begin{split}
&\textbf{Unbalanced dual}\\
\max_{\lambda,\alpha,\beta}&\ \lambda^T b + \alpha^T\mathbf{1}_{m_2} + \beta^T \mathbf{1}_{m_1}\\
s.t.&\ U + {\textstyle\sum}_k\lambda_{k} A_{k} + \mathbf{1}_{m_1}\alpha^T + \beta \mathbf{1}_{m_2}^T\leq 0\\
&\ \lambda\geq 0,\ \beta\leq 0
\end{split}
\end{equation}
\end{figure*}

\section{Experiments}
\label{sec:empirical}

In this section, we perform empirical evaluation of the proposed algorithm for ranking under constraints in the recommender system settings. We test the algorithm on recommendations of (1) movies and (2) news documents. In both cases, we use real data to train a recommender system.

\subsection{Problem setup}

We want to compute an optimized ranking of items (movies or news documents) for each user, maximizing the utility from the recommended items, while ensuring compliance with constraints, such as constraints on amount of exposure by movies/news documents across different topic areas.

To set up the optimization, we need to predict utility for all possible user-item combinations. On each data set, we train an embedding-based neural net, which predicts item utility to the user as a non-linear function of user and item embeddings. See \Cref{appendix:recommender} for training details. The predicted utility is in $[1,5]$ range for both movie and news data sets. Learned user embeddings constitute user covariates $X$. (Our proposed method uses covariates to predict $\lambda$ (see \Cref{algo:b}).) For each user, the trained recommender system outputs a vector $u\in\mathbb{R}^{m_1}$ of utilities over available items. We construct utility matrix $U$ over item-rank combinations as $U=u\gamma^T$, where $\gamma = 1/\log_2(j+1)$ for each rank $j\in 1:m_2$ ($m_1\geq m_2$), as in the discounted cumulative gain (DCG) framework \citep{jarvelin2002cumulated,singh2018fairness,biega2018equity}. $\gamma$ captures exposure from being placed in rank position $i$ (i.e., a discount on utility as we go down in the rank). All constraint matrices have the form $A_k=a_k\gamma^T$ with $a_k\in \mathbb{R}^{m_1}$. Constraints are data-specific and are discussed in more detail below. 

Given this fixed discounting problem structure, we can use the efficient $O(m_1 \log m_1)$ method for recovery of optimal assignment from the adjusted utility matrix $S=U+\sum_k\lambda_{k} A_{k}$, as discussed.\footnote{Per earlier discussion, we actually compute $S=U+\sum_k(1+\varepsilon) \lambda_{k} A_{k}$ to favor constraint-compliant solutions in case of ties. We select $\varepsilon$ that minimizes train subset constraint violation probability -- 
from the candidate set $\{0\}\cup \{i\cdot 10^{-j}\ |\ i\in 1:9,\ j\in 1:4\}$. 
} However, shadow prices $\lambda$ are unknown. We compare the following strategies for computing $\lambda$ -- in terms of speed, achieved utility, and probability of constraint compliance on a holdout set of users:
\begin{itemize}
    \item {\bf No optimization:} No accounting for constraints ($\lambda=\mathbf{0}$), showing top items by utility (benchmark).
    \item {\bf Optimal lambda:} Dual optimization program (\cref{eq:dual}) is solved for each holdout user to get $\lambda$ (time-intensive).
    \item {\bf Mean lambda:} An average shadow price vector $\bar\lambda$ across users in the train set is used to compute the ranking (i.e., an intercept-only covariate-free predictor).
    \item {\bf KNeighbors lambda:} K-nearest neighbor regressor,\footnote{We use scalable ball-tree k-nearest neighbor regressor \citep{scikit-learn}, where neighbor points are weighted by the inverse of their Euclidean distance to the point for which the prediction is made, so closer neighbors have greater influence. We use $k=10$ nearest neighbors, following \citet{nigsch2006melting}.} trained on the train set, is used to predict personalized $\hat\lambda$ from user's covariates. This is our proposed method.
\end{itemize}

Across experiments, in order to solve \cref{eq:dual}, we use CBC (COIN-OR Branch-and-Cut) solver within CVXPY library \citep{diamond2016cvxpy}, which is capable of efficiently handling a large number of constraints. 
Following \citet{zhernov2020nodehopper}, all reported running times include exclusively time spent solving the optimization problem on users in the test set and exclude time spent on data generation, reading, and pre-processing, as well as time spent on any offline computation that can be done beforehand and need not be completed live. All experiments are implemented in Python and are run on the same uniform hardware.\footnote{
All evaluations are performed on the same machine (2.7 GHz Quad-Core Intel Core i5-6400).}
\begin{figure*}[h]
\setlength{\belowcaptionskip}{0.1\baselineskip}
\centering
\begin{subfigure}{.6\textwidth}
  \centering
  \includegraphics[width=1\linewidth]{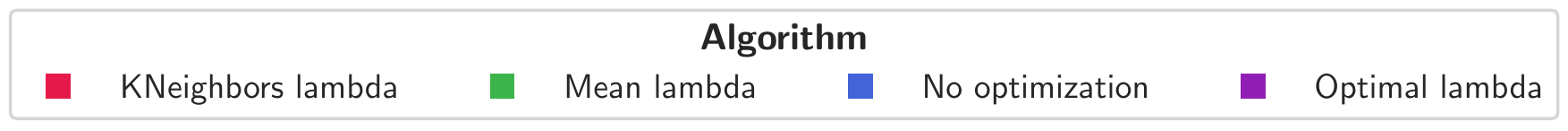}
\end{subfigure}\hspace{5mm}%
\begin{subfigure}{.25\textwidth}
  \centering
  \includegraphics[width=1\linewidth]{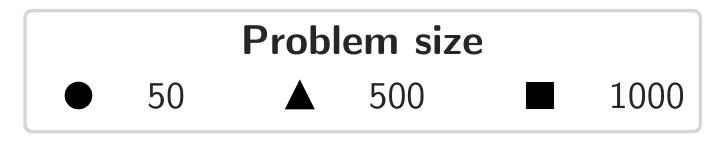}
\end{subfigure}
\begin{subfigure}{.45\textwidth}
  \centering
  \includegraphics[width=1\linewidth]{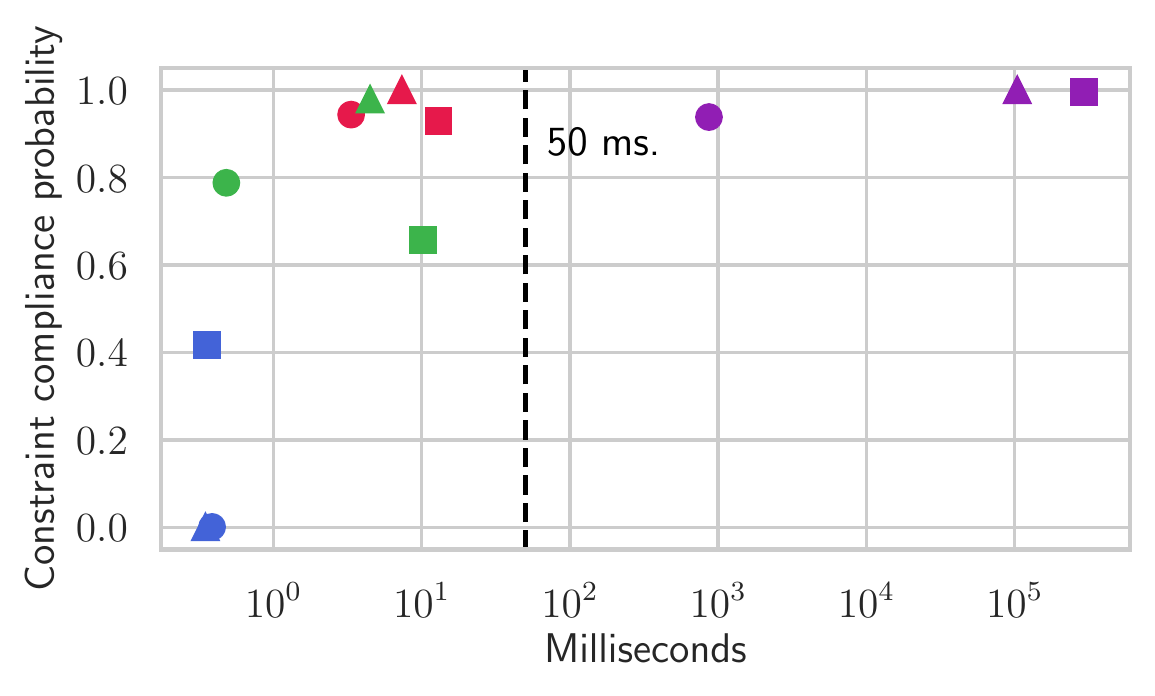}
  \caption{MovieLens data ($K=5$ constraints).}
\end{subfigure}\hspace{5mm}%
\begin{subfigure}{.45\textwidth}
  \centering
  \includegraphics[width=1\linewidth]{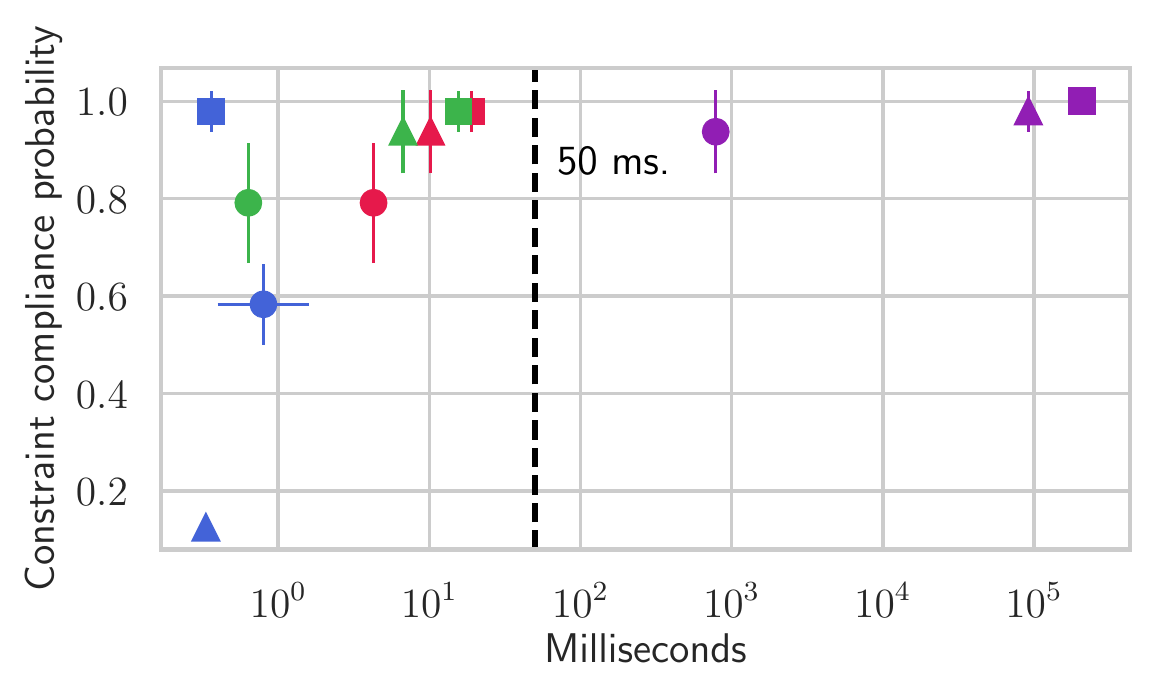}
  \caption{YOW news data ($K=8$ constraints).}
\end{subfigure}
\caption{Algorithm computation time vs. constraint compliance on MovieLens and YOW news data sets. Results are averages across holdout users in three scenarios, where we rank (a) top 50, (b) top 500, and (c) 1,000 items from among the 1,000 highest-utility items for each user, subject to constraints. Error bars indicate two standard errors of the mean ($n_\text{movielens}=250$ and $n_\text{yow}=6$ per algorithm-problem size combination; some are not visible because the error is small). Dashed line indicates 50 millisecond real-time latency requirement.}
\label{fig:performance}
\end{figure*}

\subsection{Evaluation}

We perform evaluation on MovieLens 25M\footnote{\scriptsize \url{https://grouplens.org/datasets/movielens/25m/}} and YOW news\footnote{\scriptsize \url{https://users.soe.ucsc.edu/~yiz/papers/data/YOWStudy/}} recommendation data sets.

\subsubsection{MovieLens} MovieLens data contains 25 million ratings for 62 thousand movies by 162 thousand users. There is information on how each movie scores on a variety of tags in terms of relevance. We classify each movie as being in top 5\% of movies on the following tags: (1) Gay character, (2) Race issues, (3) Freedom of speech, and (4) Science fiction. We also know the year of release for each movie. In training the recommender system, we use the full data. In optimization experiments, as one of the benchmarks, we need to solve the time-intensive full dual optimization program for each holdout user (`optimal lambda'), which is time-consuming. For that reason, we perform optimization experiments on 1,000 users sampled from the full data -- so that we can compute both scalable and time-intensive optimal solutions within a reasonable amount of time. Train set contains 750 users and holdout set contains 250 users (randomly split).

\subsubsection{YOW news} YOW news data contains evaluations of $\sim$6 thousand news documents by 24 users. Specifically, data contains $\sim$10 thousand relevance scores assigned by users, a subset of all possible user-document evaluations. The data set also contains document tags, which allow us to classify news documents into a set of topics. Based on tags, we classify each document using a set of binary indicators characterizing the topics: (1) Science and Technology, (2) Health, (3) Business, (4) Entertainment, (5) World, (6) Politics, (7) Sport, and (8) Environment. Train set contains 18 users and holdout set contains 6 users, the split being random.

\subsubsection{Scenarios} The optimization is on a per-user basis, its complexity driven by the number of items, rank positions, and constraints. We consider three optimization scenarios, where we rank (a) top 50, (b) top 500, and (c) 1,000 items from among the 1,000 highest-utility items for each user, while imposing data-specific constraints. For each user $l$, we have $u^{(l)}\in \mathbb{R}^{1000}$, $\gamma\in \mathbb{R}^{50},\mathbb{R}^{500},\mathbb{R}^{1000}$, $U^{(l)}=u^{(l)}\gamma^T$, $a^{(l)}\in \mathbb{R}^{1000}$, $A_k^{(l)} = a_k^{(l)}\gamma^T$. $a_k^{(l)}$ is a binary vector capturing whether each of 1,000 considered items for user $l$ belongs to topic $k$ (in MovieLens data, it also contains (scaled) delta of the movie release year relative to 1990). $b$ contains corresponding constraint scalars, selected to promote recommendation diversity while ensuring program feasibility. $b\in \mathbb{R}^5$ for MovieLens and $b\in \mathbb{R}^8$ for YOW news. \Cref{tab:topics} shows specific constraints. The mean per-user optimization strategy performance across scenarios and the associated confidence intervals are estimated on the holdout user set.

\begin{table*}[h]
\caption{Inequality constraints in MovieLens and YOW news experiments. Constraints on \% exposure are in terms of the total exposure: $\sum_j\gamma_j = \sum_j 1/\log_2(j+1)$. We consider three scenarios, where we rank (a) top 50, (b) top 500, and (c) 1,000 items from among 1,000 highest-utility items for each user. Constraint scalars $b_k$ differ across scenarios to ensure problem feasibility.}
\label{tab:topics}
\vskip 0.15in
\begin{subtable}[h]{1.\textwidth}
\caption{Constraints on required \% of exposure by topic and exposure-weighted movie release year in MovieLens recommendations data set.}
\vskip 0.05in
\centering
\begin{small}
\begin{sc}
\begin{tabular}{cccccc}
\toprule
No.& $\textbf{tr}(A^TP)$ & Ineq. & Top 50 $b_k$ & Top 500 $b_k$ & Top 1000 $b_k$ \\ \midrule
1&Gay character (queer) movies - \% total exposure & $\geq$& 10\% & 5\%& 1.5\%  \\ 
2&Racial issues movies - \% total exposure &$\geq$& 10\% & 5\%& 1.5\%   \\ 
3&Freedom of speech movies - \% total exposure &$\geq$&10\% & 5\%& 1.5\%   \\
4&Science fiction movies - \% total exposure &$\geq$& 10\% & 5\%& 1.5\%  \\ 
5&Average[(Movie $i$ release year - 1990)$\times \gamma_j/100$] &$\geq$& 0 & 0& 0 \\ \bottomrule
\end{tabular}
\end{sc}
\end{small}
\end{subtable}
\par\vskip 0.2in
\begin{subtable}[h]{1.\textwidth}
\caption{Topic characteristics and constraints on required \% of exposure by topic in YOW news recommendations data set.}
\vskip 0.05in
\centering
\begin{small}
\begin{sc}
\begin{tabular}{ccccccc}
\toprule
No.& $\textbf{tr}(A^TP)$ & Ineq. & Top 50 $b_k$ & Top 500 $b_k$ & Top 1000 $b_k$ & \% docs. in data \\ \midrule
1&Science and Tech. - \% total exposure & $\geq$ & 30\%& 30\%& 20\% & 15.6\% \\ 
2&Health - \% total exposure & $\geq$ & 20\%& 20\% &15\% & 9.6\%   \\ 
3&Business - \% total exposure & $\leq$ & 10\%& 10\%& 20\% & 10.1\%   \\ 
4&Entertainment - \% total exposure & $\leq$& 10\%& 10\%& 20\% & 14.1\%   \\
5&World - \% total exposure & $\leq$& 10\%& 10\%& 20\% & 15.5\%   \\ 
6&Politics - \% total exposure & $\leq$& 10\%& 10\%& 20\% & 9.2\%  \\ 
7&Sport - \% total exposure & $\leq$& 10\%& 10\%& 20\% & 3.6\%   \\ 
8&Environment - \% total exposure & $\geq$& 5\%& 5\% & 2\% & 1.9\% \\ \bottomrule
\end{tabular}
\end{sc}
\end{small}
\end{subtable}
\end{table*}

\subsection{Results} 
\Cref{fig:performance} shows the constraint compliance and computing time across different optimization strategies on MovieLens and YOW news data sets. `Optimal lambda' strategy tends to yield the best constraint compliance of all strategies, but does not meet latency requirements.\footnote{Note that the optimal strategy does not always achieve 100\% constraint compliance -- this is in line with the notion that even the optimal solution is approximate, as discussed earlier.} At the same time, we see that the proposed prediction-based approach `KNeighbors lambda' performs close to optimal in terms of constraint compliance and is within the 50 millisecond latency requirement -- even when ranking all 1,000 items under 5 or more constraints. Importantly, \citet{zhernov2020nodehopper} report inability to solve problems of such size ($\geq 500$ ranked objects and $\geq 5$ constraints) in real time. This highlights the speed advantage of prediction-based methods and suggests they are an attractive approach to solving large ranking problems in real time. `Mean lambda' method underperforms compared to `KNeighbors lambda' approach in terms of achieved constraint compliance on MovieLens data, suggesting that capturing user heterogeneity can be advantageous. `Mean lambda' and `KNeighbors lambda' achieve similar constraint compliance on YOW news data set. Under no optimization, we get the worst constraint compliance. Algorithm differences in achieved utility were small in magnitude and mostly not significant,\footnote{No optimization MovieLens utility was significantly lower.} in line with findings that the price of imposing diversity constraints is often low \citep{bandi2021price}. See \Cref{appendix:algperf} for regression output.

\section{Related work}

Many methods exist for ranking under multiple objectives \citep{singh2018fairness,yang2017measuring,zehlike2017fa,celis2017ranking,asudeh2019designing,biega2018equity,radlinski2008learning}. A common approach, called ranking under constraints, is to consider alternative objectives as constraints on the primary objective.

\citet{singh2018fairness,biega2018equity} have considered formulation of the ranking under constraints in recommender setting as a weighted bipartite matching program, which can handle varied constraints and flexible preferences over arbitrary item-ranking combinations. However, their proposed linear program formulation is too computationally inefficient to meet the strict latency requirement for live deployment. 
\citet{zhernov2020nodehopper} have constructed a dedicated solver to speed up solution of such a linear program, when it has a special structure. However, they have reported that their approach cannot meet requirements of real-time performance for large decision problems (for example, $\geq 500$ ranked objects together with $\geq 5$ constraints), whereas our proposed algorithm meets the 50 millisecond latency requirement when solving problems of similar and larger sizes, as we have demonstrated. Their algorithm also breaks down when a special structure of fixed discounting along the ranks is absent, whereas our algorithm offers speed-ups even in the absence of such structure.

Others have proposed algorithms that may be faster, but are not general enough to handle flexible preferences and varied constraints: \citet{asudeh2019designing} require ranking function adjustment by the user; \citet{celis2017ranking,zehlike2017fa} deal with top-n style ranking / constraints; \citet{yang2017measuring} restrict the problem to a narrow set of specific fairness measures. 

\citet{shah2017online} have shown that the dual formulation of the weighted bipartite matching problem can be first solved offline on a statistical sample of users -- to estimate `average' $\hat\lambda$ based on aggregate $U$ and $A_k$ for that sample. These shadow prices can then be used to arrive at a ranking in online settings, without having to again solve the linear program, offering a substantial speed advantage. Note that their solution does not guarantee the constraints are met on per-user basis, but instead only on average across many users. The idea of using sampling-based estimates of shadow prices for online matching has also been discussed in earlier works \citep{mehta2012online}. 
Our idea of replacing optimization with prediction builds on and generalizes this stream of research. Specifically, instead of learning a single set of shadow prices for a user population, we propose training a model to predict personalized optimal shadow prices based on user covariates, capturing user heterogeneity.

\section{Conclusion and future research directions}

In this work, we propose a scalable algorithm for ranking under constraints, based on a dual formulation of the weighted bipartite matching program. Specifically, we propose to replace online optimization with prediction. We solve for optimal shadow prices on a sample of users in offline settings, where speed is not critical. We then train a model to predict users' optimal shadow prices from users' characteristics. In online settings, the model can predict the shadow prices based on observed covariates, without solving the time-consuming optimization problem, allowing us to quickly compute the ranking for any such user. We show empirically that the proposed approximate solution to the ranking problem leads to a substantial reduction in required computing resources, meeting real-time 50 millisecond latency requirement, without much sacrifice in constraint compliance and achieved utility, allowing us to solve larger constrained ranking problems real-time than previously reported ($\geq 500$ ranked objects and $\geq 5$ constraints) \citep{zhernov2020nodehopper}. Our method thus enables the deployment of the constrained ranking to new large-scale problems, where latency matters. We elucidate the role of rearrangement inequality / Monge problem structure in achieving the speed-ups. 

{\bfseries Future directions\ } There is a place for more research on fast unconstrained optimal assignment methods (\cref{eq:primal_unc}), which our proposed algorithm depends on when computing ranking from the adjusted utility matrix $S$. 
The structuring of the problem in terms of Monge matrices, a generalization of rearrangement inequalities, speeds up the solution. 
However, 
most research has assumed arrays are precisely Monge. This condition may be hard to satisfy, e.g., because of noise. Efficient statistical estimation of the best Monge approximation to a matrix that is not strictly Monge \citep{hutter2020estimation} could enable faster approximate ranking on a larger problem set, warranting more research. Outside of decision problems with Monge structure, there have been promising results on approximating the Hungarian algorithm solution with neural net predictions \citep{lee2018deep}. This deep learning approach to  solving combinatorial optimization problems deserves further attention. In general, the prediction of optimization solutions as a way to reduce computation time appears to be a promising research direction.

\section*{Acknowledgements}

This work was supported The Sanford C. Bernstein \& Co. Center for Leadership and Ethics at Columbia Business School.

\bibliography{bibliography}

\begin{thebibliography}{53}
\providecommand{\natexlab}[1]{#1}
\providecommand{\url}[1]{\texttt{#1}}
\expandafter\ifx\csname urlstyle\endcsname\relax
  \providecommand{\doi}[1]{doi: #1}\else
  \providecommand{\doi}{doi: \begingroup \urlstyle{rm}\Url}\fi

\bibitem[Aggarwal et~al.(1992)Aggarwal, Bar-Noy, Khuller, Kravets, and
  Schieber]{aggarwal1992efficient}
Aggarwal, A., Bar-Noy, A., Khuller, S., Kravets, D., and Schieber, B.
\newblock Efficient minimum cost matching using quadrangle inequality.
\newblock In \emph{STACS}, pp.\  583--592. IEEE Computer Society, 1992.

\bibitem[Ansari et~al.(2000)Ansari, Essegaier, and Kohli]{ansari2000internet}
Ansari, A., Essegaier, S., and Kohli, R.
\newblock Internet recommendation systems.
\newblock \emph{Journal of Marketing Research}, 37\penalty0 (3), 2000.

\bibitem[Asudeh et~al.(2019)Asudeh, Jagadish, Stoyanovich, and
  Das]{asudeh2019designing}
Asudeh, A., Jagadish, H., Stoyanovich, J., and Das, G.
\newblock Designing fair ranking schemes.
\newblock In \emph{{ACM SIGMOD}}, pp.\  1259--1276, 2019.

\bibitem[Avis(1983)]{avis1983survey}
Avis, D.
\newblock A survey of heuristics for the weighted matching problem.
\newblock \emph{Networks}, 13\penalty0 (4):\penalty0 475--493, 1983.

\bibitem[Bandi \& Bertsimas(2021)Bandi and Bertsimas]{bandi2021price}
Bandi, H. and Bertsimas, D.
\newblock The price of diversity.
\newblock \emph{arXiv preprint arXiv:2107.03900}, 2021.

\bibitem[Biega et~al.(2018)Biega, Gummadi, and Weikum]{biega2018equity}
Biega, A.~J., Gummadi, K.~P., and Weikum, G.
\newblock Equity of attention: Amortizing individual fairness in rankings.
\newblock In \emph{ACM SIGIR}, pp.\  405--414, 2018.

\bibitem[Birkhoff(1940)]{birkhoff1940lattice}
Birkhoff, G.
\newblock \emph{Lattice theory}, volume~25.
\newblock American Mathematical Soc., 1940.

\bibitem[Bougleux et~al.(2017)Bougleux, Ga{\"u}z{\`e}re, and
  Brun]{bougleux2017hungarian}
Bougleux, S., Ga{\"u}z{\`e}re, B., and Brun, L.
\newblock A {Hungarian} algorithm for error-correcting graph matching.
\newblock In \emph{GbRPR}, pp.\  118--127. Springer, 2017.

\bibitem[Boyd \& Vandenberghe(2004)Boyd and Vandenberghe]{boyd2004convex}
Boyd, S. and Vandenberghe, L.
\newblock \emph{Convex optimization}.
\newblock Cambridge University Press, 2004.

\bibitem[Bubeck(2014)]{bubeck2014convex}
Bubeck, S.
\newblock Convex optimization: Algorithms and complexity.
\newblock \emph{arXiv preprint arXiv:1405.4980}, 2014.

\bibitem[Burkard et~al.(2012)Burkard, Dell'Amico, and
  Martello]{burkard2012assignment}
Burkard, R., Dell'Amico, M., and Martello, S.
\newblock \emph{Assignment problems: Revised reprint}.
\newblock SIAM, 2012.

\bibitem[Burkard(2007)]{burkard2007monge}
Burkard, R.~E.
\newblock Monge properties, discrete convexity and applications.
\newblock \emph{European Journal of Operational Research}, 176\penalty0
  (1):\penalty0 1--14, 2007.

\bibitem[Burkard et~al.(1996)Burkard, Klinz, and
  Rudolf]{burkard1996perspectives}
Burkard, R.~E., Klinz, B., and Rudolf, R.
\newblock Perspectives of {Monge} properties in optimization.
\newblock \emph{Discrete Applied Mathematics}, 70\penalty0 (2):\penalty0
  95--161, 1996.

\bibitem[Celis et~al.(2017)Celis, Straszak, and Vishnoi]{celis2017ranking}
Celis, L.~E., Straszak, D., and Vishnoi, N.~K.
\newblock Ranking with fairness constraints.
\newblock \emph{arXiv preprint arXiv:1704.06840}, 2017.

\bibitem[Chade et~al.(2017)Chade, Eeckhout, and Smith]{chade2017sorting}
Chade, H., Eeckhout, J., and Smith, L.
\newblock Sorting through search and matching models in economics.
\newblock \emph{Journal of Economic Literature}, 55\penalty0 (2):\penalty0
  493--544, 2017.

\bibitem[Covington et~al.(2016)Covington, Adams, and Sargin]{covington2016deep}
Covington, P., Adams, J., and Sargin, E.
\newblock Deep neural networks for {YouTube} recommendations.
\newblock In \emph{ACM RECSYS}, pp.\  191--198, 2016.

\bibitem[Derigs et~al.(1986)Derigs, Goecke, and Schrader]{derigs1986monge}
Derigs, U., Goecke, O., and Schrader, R.
\newblock Monge sequences and a simple assignment algorithm.
\newblock \emph{Discrete Applied Mathematics}, 15\penalty0 (2-3):\penalty0
  241--248, 1986.

\bibitem[Devroye et~al.(1994)Devroye, Gyorfi, Krzyzak, and
  Lugosi]{devroye1994strong}
Devroye, L., Gyorfi, L., Krzyzak, A., and Lugosi, G.
\newblock On the strong universal consistency of nearest neighbor regression
  function estimates.
\newblock \emph{The Annals of Statistics}, pp.\  1371--1385, 1994.

\bibitem[Diamond \& Boyd(2016)Diamond and Boyd]{diamond2016cvxpy}
Diamond, S. and Boyd, S.
\newblock {CVXPY}: {A} {P}ython-embedded modeling language for convex
  optimization.
\newblock \emph{Journal of Machine Learning Research}, 17\penalty0
  (83):\penalty0 1--5, 2016.

\bibitem[Duff \& Koster(2001)Duff and Koster]{duff2001algorithms}
Duff, I.~S. and Koster, J.
\newblock On algorithms for permuting large entries to the diagonal of a sparse
  matrix.
\newblock \emph{SIAM Journal on Matrix Analysis and Applications}, 22\penalty0
  (4):\penalty0 973--996, 2001.

\bibitem[Dufoss{\'e} \& U{\c{c}}ar(2016)Dufoss{\'e} and
  U{\c{c}}ar]{dufosse2016notes}
Dufoss{\'e}, F. and U{\c{c}}ar, B.
\newblock Notes on {Birkhoff--von Neumann} decomposition of doubly stochastic
  matrices.
\newblock \emph{Linear Algebra and its Applications}, 497:\penalty0 108--115,
  2016.

\bibitem[Edmonds \& Karp(1972)Edmonds and Karp]{edmonds1972theoretical}
Edmonds, J. and Karp, R.~M.
\newblock Theoretical improvements in algorithmic efficiency for network flow
  problems.
\newblock \emph{Journal of the ACM}, 19\penalty0 (2):\penalty0 248--264, 1972.

\bibitem[Fortin \& Rudolf(1998)Fortin and Rudolf]{fortin1998weak}
Fortin, D. and Rudolf, R.
\newblock Weak {Monge} arrays in higher dimensions.
\newblock \emph{Discrete Mathematics}, 189\penalty0 (1-3):\penalty0 105--115,
  1998.

\bibitem[Goodfellow et~al.(2016)Goodfellow, Bengio, and
  Courville]{Goodfellow-et-al-2016}
Goodfellow, I., Bengio, Y., and Courville, A.
\newblock \emph{Deep Learning}.
\newblock MIT Press, 2016.

\bibitem[Gusfield(1992)]{eff_alg}
Gusfield, D.
\newblock \emph{Handbooks in Operations Research and Management Science},
  volume~3, chapter 8: Design (with Analysis) of Efficient Algorithms.
\newblock Elsevier, 1992.

\bibitem[Hardy et~al.(1952)Hardy, Littlewood, and Pólya]{hardy1952}
Hardy, G.~H., Littlewood, J.~E., and Pólya, G.
\newblock \emph{Inequalities}.
\newblock Cambridge University Press, 1952.

\bibitem[Holstermann(2017)]{holstermann2017generalization}
Holstermann, J.
\newblock A generalization of the rearrangement inequality.
\newblock \emph{Mathematical Reflections}, 5:\penalty0 503--507, 2017.

\bibitem[H{\"u}tter et~al.(2020)H{\"u}tter, Mao, Rigollet, Robeva,
  et~al.]{hutter2020estimation}
H{\"u}tter, J.-C., Mao, C., Rigollet, P., Robeva, E., et~al.
\newblock Estimation of {Monge} matrices.
\newblock \emph{Bernoulli}, 26\penalty0 (4):\penalty0 3051--3080, 2020.

\bibitem[J{\"a}rvelin \& Kek{\"a}l{\"a}inen(2002)J{\"a}rvelin and
  Kek{\"a}l{\"a}inen]{jarvelin2002cumulated}
J{\"a}rvelin, K. and Kek{\"a}l{\"a}inen, J.
\newblock Cumulated gain-based evaluation of {IR} techniques.
\newblock \emph{ACM TOIS}, 20\penalty0 (4):\penalty0 422--446, 2002.

\bibitem[Kingma \& Ba(2014)Kingma and Ba]{kingma2014adam}
Kingma, D.~P. and Ba, J.
\newblock Adam: A method for stochastic optimization.
\newblock \emph{arXiv preprint arXiv:1412.6980}, 2014.

\bibitem[Kuhn(1955)]{kuhn1955hungarian}
Kuhn, H.~W.
\newblock The {Hungarian} method for the assignment problem.
\newblock \emph{Naval Research Logistics Quarterly}, 2\penalty0 (1-2):\penalty0
  83--97, 1955.

\bibitem[Kuhn(1956)]{kuhn1956variants}
Kuhn, H.~W.
\newblock Variants of the {Hungarian} method for assignment problems.
\newblock \emph{Naval Research Logistics Quarterly}, 3\penalty0 (4):\penalty0
  253--258, 1956.

\bibitem[Lee et~al.(2018)Lee, Xiong, Yu, and Li]{lee2018deep}
Lee, M., Xiong, Y., Yu, G., and Li, G.~Y.
\newblock Deep neural networks for linear sum assignment problems.
\newblock \emph{IEEE Wireless Communications Letters}, 7\penalty0 (6):\penalty0
  962--965, 2018.

\bibitem[Mehta(2012)]{mehta2012online}
Mehta, A.
\newblock Online matching and ad allocation.
\newblock \emph{Theoretical Computer Science}, 8\penalty0 (4):\penalty0
  265--368, 2012.

\bibitem[Miller(1968)]{miller1968response}
Miller, R.~B.
\newblock Response time in man-computer conversational transactions.
\newblock In \emph{Proceedings of the Fall Joint Computer Conference}, pp.\
  267--277, 1968.

\bibitem[Mnih \& Salakhutdinov(2008)Mnih and
  Salakhutdinov]{mnih2008probabilistic}
Mnih, A. and Salakhutdinov, R.~R.
\newblock Probabilistic matrix factorization.
\newblock In \emph{NeurIPS}, pp.\  1257--1264, 2008.

\bibitem[Nigsch et~al.(2006)Nigsch, Bender, van Buuren, Tissen, Nigsch, and
  Mitchell]{nigsch2006melting}
Nigsch, F., Bender, A., van Buuren, B., Tissen, J., Nigsch, E., and Mitchell,
  J.~B.
\newblock Melting point prediction employing k-nearest neighbor algorithms and
  genetic parameter optimization.
\newblock \emph{Journal of Chemical Information and Modeling}, 46\penalty0
  (6):\penalty0 2412--2422, 2006.

\bibitem[Pedregosa et~al.(2011)Pedregosa, Varoquaux, Gramfort, Michel, Thirion,
  Grisel, Blondel, Prettenhofer, Weiss, Dubourg, Vanderplas, Passos,
  Cournapeau, Brucher, Perrot, and Duchesnay]{scikit-learn}
Pedregosa, F., Varoquaux, G., Gramfort, A., Michel, V., Thirion, B., Grisel,
  O., Blondel, M., Prettenhofer, P., Weiss, R., Dubourg, V., Vanderplas, J.,
  Passos, A., Cournapeau, D., Brucher, M., Perrot, M., and Duchesnay, E.
\newblock Scikit-learn: Machine learning in {P}ython.
\newblock \emph{Journal of Machine Learning Research}, 12:\penalty0 2825--2830,
  2011.

\bibitem[Preis(1999)]{preis1999linear}
Preis, R.
\newblock Linear time 1/2-approximation algorithm for maximum weighted matching
  in general graphs.
\newblock In \emph{STACS}, pp.\  259--269. Springer, 1999.

\bibitem[Radlinski et~al.(2008)Radlinski, Kleinberg, and
  Joachims]{radlinski2008learning}
Radlinski, F., Kleinberg, R., and Joachims, T.
\newblock Learning diverse rankings with multi-armed bandits.
\newblock In \emph{ICML}, pp.\  784--791, 2008.

\bibitem[Robertson(1977)]{robertson1977probability}
Robertson, S.~E.
\newblock The probability ranking principle in {IR}.
\newblock \emph{Journal of Documentation}, 1977.

\bibitem[Roth et~al.(1993)Roth, Rothblum, and Vande~Vate]{roth1993stable}
Roth, A.~E., Rothblum, U.~G., and Vande~Vate, J.~H.
\newblock Stable matchings, optimal assignments, and linear programming.
\newblock \emph{Mathematics of Operations Research}, 18\penalty0 (4):\penalty0
  803--828, 1993.

\bibitem[Schrijver(2003)]{schrijver2003combinatorial}
Schrijver, A.
\newblock \emph{Combinatorial optimization: Polyhedra and efficiency},
  volume~24.
\newblock Springer Science \& Business Media, 2003.

\bibitem[Sethumadhavan(2009)]{sethumadhavan2009survey}
Sethumadhavan, S.
\newblock \emph{A survey of Monge properties}.
\newblock PhD thesis, Cochin University of Science and Technology, India, 2009.

\bibitem[Shah et~al.(2017)Shah, Soni, and Chevalier]{shah2017online}
Shah, P., Soni, A., and Chevalier, T.
\newblock Online ranking with constraints: A primal-dual algorithm and
  applications to web traffic-shaping.
\newblock In \emph{ACM SIGKDD}, pp.\  405--414. ACM, 2017.

\bibitem[Singh \& Joachims(2018)Singh and Joachims]{singh2018fairness}
Singh, A. and Joachims, T.
\newblock Fairness of exposure in rankings.
\newblock In \emph{ACM SIGKDD}, pp.\  2219--2228. ACM, 2018.

\bibitem[Tomizawa(1971)]{tomizawa1971some}
Tomizawa, N.
\newblock On some techniques useful for solution of transportation network
  problems.
\newblock \emph{Networks}, 1\penalty0 (2):\penalty0 173--194, 1971.

\bibitem[Vaidyanathan(2013)]{vaidyanathan2013faster}
Vaidyanathan, B.
\newblock Faster strongly polynomial algorithms for the unbalanced
  transportation problem and assignment problem with monge costs.
\newblock \emph{Networks}, 62\penalty0 (2):\penalty0 136--148, 2013.

\bibitem[Vince(1990)]{vince1990rearrangement}
Vince, A.
\newblock A rearrangement inequality and the permutahedron.
\newblock \emph{The American Mathematical Monthly}, 97\penalty0 (4):\penalty0
  319--323, 1990.

\bibitem[White(2000)]{white2000requirement}
White, D.~M.
\newblock The requirement of race-conscious evaluation of {LSAT} scores for
  equitable law school admissions.
\newblock \emph{Berkeley La Raza LJ}, 12:\penalty0 399, 2000.

\bibitem[Yang \& Stoyanovich(2017)Yang and Stoyanovich]{yang2017measuring}
Yang, K. and Stoyanovich, J.
\newblock Measuring fairness in ranked outputs.
\newblock In \emph{ACM SSDBM}, pp.\  1--6, 2017.

\bibitem[Zehlike et~al.(2017)Zehlike, Bonchi, Castillo, Hajian, Megahed, and
  Baeza-Yates]{zehlike2017fa}
Zehlike, M., Bonchi, F., Castillo, C., Hajian, S., Megahed, M., and
  Baeza-Yates, R.
\newblock {FA*IR}: A fair top-k ranking algorithm.
\newblock In \emph{ACM CIKM}, pp.\  1569--1578, 2017.

\bibitem[Zhernov et~al.(2020)Zhernov, Dvijotham, Lobov, Calian, Gong,
  Chandrashekar, and Mann]{zhernov2020nodehopper}
Zhernov, A., Dvijotham, K.~D., Lobov, I., Calian, D.~A., Gong, M.,
  Chandrashekar, N., and Mann, T.~A.
\newblock The {NodeHopper}: Enabling low latency ranking with constraints via a
  fast dual solver.
\newblock In \emph{ACM SIGKDD}, pp.\  1285--1294, 2020.

\end{thebibliography}
\bibliographystyle{icml2022}

\appendix

\section{Problem structure for further speed up}
\label{appendix:a}

We now discuss generalization of and alternatives to fixed discounting problem structure for speeding up maximum weight assignment on the adjusted utility matrix $S$. 

\subsection{Monge structure}
Rearrangement inequalities have been extended from a product of sequence elements in \cref{eq:rear} to supermodular functions of variable pairs \citep{holstermann2017generalization,vince1990rearrangement,chade2017sorting}, where identity permutation still guarantees optimal assignment. 
In terms of matrices that we work with, this generalization is equivalent to an observation that if $S$ is an {\it (inverse) Monge} matrix, then optimal permutation matrix $P$ does not depend on $S$ and has the form of an identity matrix $P=I_{m\times m}$ \citep{sethumadhavan2009survey,burkard1996perspectives,burkard2012assignment}. A matrix $U$ is called inverse Monge when for $1\leq i_1< i_2\leq m$ and $1\leq j_1< j_2\leq m$: $U[i_1,j_1]+U[i_2,j_2]\geq U[i_1,j_2]+U[i_2,j_1]\text{.}$
For $u$ and $\gamma$ non-increasing, $U = u\gamma^T$ is inverse Monge \citep{burkard2007monge}.\footnote{For $1\leq i_1\leq i_2\leq m$ and $1\leq j_1\leq j_2\leq m$, let $u[i_1]\geq u[i_2]$, $\gamma[j_1] \geq \gamma[j_2]$.  $U = u\gamma^T$. Inverse Monge property $u[i_1] \gamma[j_1] + u[i_2] \gamma[j_2] \geq u[i_1] \gamma[j_2] + u[i_2] \gamma[j_1]$ is equivalent to $(u[i_1]-u[i_2])(\gamma[j_1]-\gamma[j_2])\geq 0$, which always holds because $u[i_1]\geq u[i_2]$ and $\gamma[j_1]-\gamma[j_2]\geq 0$. Thus, $U$ is inverse Monge.}
What conditions guarantee $S=U+\sum_k\lambda_{k} A_{k}$ ($\lambda_k\geq0$) is inverse Monge too? As we have shown, fixed discounting $S=(u+\sum_k\lambda_ka_k)\gamma^T$ ensures that $S$, after sorting on the first column, is represented as a dot product of two non-increasing vectors and is thus inverse Monge, so identity permutation extracts an optimal assignment. Matrix that is Monge after sorting is called a permuted Monge matrix \citep{hutter2020estimation}.

When fixed discounting is not available as a decision problem structure, we still have some options. An important property of inverse Monge matrices is that they are closed under several operations \citep{burkard1996perspectives}. Consider two inverse Monge matrices $C,D\in \mathbb{R}^{m\times m}$ and two vectors $\alpha,\beta \in \mathbb{R}^{m}$. Then the following matrices are inverse Monge as well: (1) transpose $C^T$; (2) $\tau C$ for $\tau\geq0$; (3) sum $C+D$; and (4) matrix $F$, where $F_{ij}=C_{ij}+\alpha_i+\beta_j$. 
Properties (2) and (3) imply that a linear combination with non-negative coefficients of inverse Monge matrices is itself an inverse Monge array. Thus, if $U$ and all $A_k$ are inverse Monge, then, because $\lambda_k\geq 0$, $S$ will also be inverse Monge, so optimal assignment problem could be obtained as an identity permutation ($O(m)$ worst time complexity). These properties enable efficient solution of additional constrained ranking problems. For example, this framework accommodates utility and/or constraint matrices of $\alpha_i+\beta_j$ variety for arbitrary $\alpha,\beta \in \mathbb{R}^{m}$, which do not fall within the fixed discounting framework prevalent in the literature. When Monge structure is not available, we can check if the matrix is weak Monge \citep{fortin1998weak,derigs1986monge,burkard1996perspectives} -- the most general condition we are aware of sufficient for the identity permutation to yield an optimal assignment. 

\subsection{Other approaches}
Without Monge condition, we can use the greedy $1/2$-approximation algorithm \citep{avis1983survey,preis1999linear,eff_alg} with worst-case complexity $O(m^2)$ before using the $O(m^3)$ Hungarian algorithm. However, as mentioned earlier, the weight of its matching can be as low as $1/2$ of the maximum weight computed by the Hungarian algorithm. There exist specialized conditions, known as box inequalities, which are related to the Monge property, such that when the matching weight matrix satisfies these conditions, the assignment obtained via the greedy algorithm is in fact optimal -- these have been successfully used to address the shortest superstring problem in biology \citep{eff_alg,burkard2012assignment}. However, their specialized form makes them non-trivial to apply in typical ranking-under-constraints settings. There is also research on using neural nets to approximate Hungarian algorithm optimal assignment -- this approach could be attempted if no special structure in the problem is present, but a fast solution with complexity $O(m^2)$ or $O(m)$ (which depends on the neural net type) is desirable \citep{lee2018deep}.

\section{Recommender system training}
\label{appendix:recommender}

Recommender systems that we train on the movie and news recommendation data sets to predict user-item utility are of the matrix factorization flavor \citep{mnih2008probabilistic}, where the dot product operation from the traditional matrix factorization framework is replaced with a neural net computation, similar to \citet{covington2016deep}. Each user and each item in the data set is assigned an embedding vector of dimension $20$ (user embeddings are denoted $e_u$ and item embeddings are denoted $e_v$). Additionally, each user and each item is assigned a vector of dimension $5$, where each element is an intercept term for each of the observed rating/relevance values: $\{1,2,3,4,5\}$ (user and item intercepts are, respectively, denoted $g_u$ and $g_v$). 

Based on these embedding and intercept vectors, for a given user-item combination, the deep net outputs the probabilities for each rating level. First, user and item embeddings are concatenated into a single vector and then processed by a neural net with a hidden layer of dimension $15$, followed by rectified nonlinearity and dropout of $0.1$ \citep{Goodfellow-et-al-2016}, outputting $5$ deterministic utilities corresponding to different rating values. User and item intercept values are added to these deterministic utilities, for the corresponding ratings. 
Probabilities are obtained by passing the deterministic utilities through a softmax transformation. Point prediction is obtained as a probability-weighted sum of $\{1,2,3,4,5\}$ rating values.

The parameters of the neural net are optimized using Adam mini batch gradient descent algorithm \citep{kingma2014adam} with learning rate $0.01$. We train the model for 5 epochs (5 passes through all training data) with a mini batch of size $200$ (each iteration, a gradient update is computed based on $200$ sampled observations). When training the net, we use cross entropy loss based on predicted rating probabilities and observed ratings to compute the gradients. 

\section{Algorithm performance}
\label{appendix:algperf}

\Cref{apd:tab1,apd:tab2} show analysis of algorithm performance.

\begin{table*}[ht] \centering
  \caption{Regression analysis of algorithm performance: MovieLens data ($K=5$ constraints).}
  \label{apd:tab1}
\vskip 0.15in
\begin{small}
\begin{sc}
\begin{tabular}{lccc}
\toprule
& \multicolumn{3}{c}{Performance metric}
 \\ \cmidrule{2-4}
& \multicolumn{1}{c}{Log10 comp. time (ms.)} & \multicolumn{1}{c}{Constraint compl. prob.} & \multicolumn{1}{c}{Utility}  \\ \midrule 
Optimal lambda (vs. KNeighbors)  & 3.64$^{***}$ & 0.02$^{***}$ & -0.00$^{}$ \\
  & (3.59 , 3.69) & (0.01 , 0.03) & (-0.79 , 0.78) \\
Mean lambda (vs. KNeighbors) & -0.39$^{***}$ & -0.15$^{***}$ & -0.00$^{}$ \\
  & (-0.41 , -0.37) & (-0.16 , -0.14) & (-0.79 , 0.78) \\
No optimization (vs. KNeighbors) & -1.28$^{***}$ & -0.82$^{***}$ & 0.86$^{**}$ \\
  & (-1.31 , -1.24) & (-0.83 , -0.80) & (0.08 , 1.65) \\
Problem size: 1000 (vs. 50) & 1.10$^{***}$ & 0.08$^{***}$ & 442.77$^{***}$ \\
  & (1.06 , 1.14) & (0.07 , 0.10) & (442.05 , 443.50) \\
Problem size: 500 (vs. 50) & 0.84$^{***}$ & 0.08$^{***}$ & 234.23$^{***}$ \\
  & (0.80 , 0.88) & (0.07 , 0.09) & (233.82 , 234.63) \\
Intercept & 0.19$^{***}$ & 0.90$^{***}$ & 54.75$^{***}$ \\
  & (0.16 , 0.22) & (0.89 , 0.92) & (54.26 , 55.23) \\ \midrule
 Observations & 3,000 & 3,000 & 3,000 \\
 $R^2$ & 0.95 & 0.86 & 1.00 \\
 Adjusted $R^2$ & 0.95 & 0.86 & 1.00 \\
 \bottomrule \\[-1.8ex]
Note: & \multicolumn{3}{r}{$^{*}$p$<$0.1; $^{**}$p$<$0.05; $^{***}$p$<$0.01} \\
 & \multicolumn{3}{r}{Standard errors are heteroscedasticity robust (HC3)} \\
\end{tabular}
\end{sc}
\end{small}
\end{table*}

\begin{table*}[ht] \centering
  \caption{Regression analysis of algorithm performance: YOW News data ($K=8$ constraints).}
\label{apd:tab2}
\vskip 0.15in
\begin{small}
\begin{sc}
\begin{tabular}{lccc}
\toprule
& \multicolumn{3}{c}{Performance metric}
 \\ \cmidrule{2-4}
& \multicolumn{1}{c}{Log10 comp. time (ms.)} & \multicolumn{1}{c}{Constraint compl. prob.} & \multicolumn{1}{c}{Utility} \\ \midrule 
Optimal lambda (vs. KNeighbors) & 3.42$^{***}$ & 0.07$^{}$ & 0.17$^{}$ \\
  & (3.09 , 3.75) & (-0.02 , 0.16) & (-7.55 , 7.89) \\
Mean lambda (vs. KNeighbors) & -0.37$^{***}$ & -0.00$^{}$ & 0.02$^{}$ \\
  & (-0.49 , -0.25) & (-0.10 , 0.10) & (-7.72 , 7.76) \\
No optimization (vs. KNeighbors) & -1.31$^{***}$ & -0.34$^{***}$ & 0.38$^{}$ \\
  & (-1.65 , -0.98) & (-0.49 , -0.19) & (-7.36 , 8.11) \\
Problem size: 1000 (vs. 50) & 1.03$^{***}$ & 0.21$^{***}$ & 435.53$^{***}$ \\
  & (0.70 , 1.36) & (0.12 , 0.30) & (428.57 , 442.49) \\
Problem size: 500 (vs. 50) & 0.77$^{***}$ & -0.03$^{}$ & 231.51$^{***}$ \\
  & (0.46 , 1.08) & (-0.14 , 0.08) & (227.06 , 235.95) \\
Intercept & 0.37$^{***}$ & 0.84$^{***}$ & 55.70$^{***}$ \\
  & (0.14 , 0.60) & (0.75 , 0.93) & (50.85 , 60.56) \\ \midrule
 Observations & 72 & 72 & 72 \\
 $R^2$ & 0.94 & 0.56 & 1.00 \\
 Adjusted $R^2$ & 0.94 & 0.52 & 1.00 \\
\bottomrule \\[-1.8ex]
Note: & \multicolumn{3}{r}{$^{*}$p$<$0.1; $^{**}$p$<$0.05; $^{***}$p$<$0.01} \\
 & \multicolumn{3}{r}{Standard errors are heteroscedasticity robust (HC3)} \\
\end{tabular}
\end{sc}
\end{small}
\end{table*}

\end{document}